\documentstyle[psfig,draft,aps]{revtex}
\begin{document}
\draft
\title{The Strong Isospin-Breaking Correction for the Gluonic Penguin
Contribution to $\epsilon^{\prime}/
\epsilon$ at Next-to-Leading Order in the Chiral Expansion}
\author{Carl E. Wolfe\thanks{e-mail: wolfe@niobe.iucf.indiana.edu}}
\address{Nuclear Theory Center, 
Indiana University, Bloomington, IN, 47408}
\author{Kim Maltman\thanks{e-mail: maltman@fewbody.phys.yorku.ca}}
\address{Department of Mathematics and Statistics, York University, \\
4700 Keele St., Toronto, Ontario, Canada M3J 1P3 \\ and }
\address{Special Research Centre for the Subatomic Structure of Matter, \\
University of Adelaide, Australia 5005.}
\maketitle
\begin{abstract}
The strong isospin-breaking correction, $\Omega_{st}$, 
which appears in estimates of the Standard Model
value for the direct CP-violating ratio $\epsilon^{\prime}/
\epsilon$, is evaluated to next-to-leading order (NLO)
in the chiral expansion using Chiral Perturbation Theory.
The relevant linear combinations of the
unknown NLO CP-odd weak low-energy constants (LEC's) which,
in combination with 1-loop and strong LEC contributions,
are required for a complete determination at this order,
are estimated using two different models.
It is found that, to NLO, $\Omega_{st}=0.08\pm 0.05$,
significantly reduced from the
``standard'' value, $0.25\pm 0.08$, employed in recent
analyses.  The potentially significant numerical impact
of this decrease on Standard Model predictions for
$\epsilon^{\prime}/ \epsilon$, associated with the decreased
cancellation between gluonic penguin and electroweak
penguin contributions, is also discussed.
\end{abstract}

\section{Introduction}

The recent improved experimental results for the ratio of direct to indirect 
CP-violation parameters, 
$\epsilon^{\prime}/\epsilon$, obtained by both the KTEV and 
NA48 collaborations~\cite{KTeV99,na4899},
have spurred on continuing efforts to reduce the theoretical uncertainties 
in
our expectations for the value of $\epsilon^{\prime}/\epsilon$
in the Standard Model.
While short-distance effects are under control (the 
Wilson coefficients of the effective weak Hamiltonian being 
known to two-loop order)\cite{BJL93}, 
there remains significant uncertainty in the theoretical calculation of 
the long-distance $K\rightarrow\pi\pi$ hadronic matrix 
elements.  These have been estimated 
using a number of techniques and models in 
Refs.~\cite{oneoverNc,munich,lattice,lund,trieste95,trieste98_1,trieste_misc}
(for recent reviews, see Refs.~\cite{reviews,jamin99}).

An important ingredient in the calculation of the hadronic matrix elements
is the inclusion of isospin-breaking (IB) effects.
Strong isospin breaking arising from 
the up/down quark mass difference, $\delta m\equiv m_d-m_u$, induces a 
$\Delta I = 3/2$ contribution to the $K\rightarrow\pi\pi$
matrix elements of the
gluonic penguin operator, $Q_6${\begin{footnote}
{We employ throughout the paper the notation of 
Ref.~\cite{BJL93}
for the four-quark operators of the effective weak Hamiltonian.}
\end{footnote}},
which, in the isospin limit, is pure $\Delta I=1/2$.
This `leakage' of octet ($\Delta I = 1/2$) strength into 
the $\Delta I = 3/2$ component of the $K^0\to\pi\pi$ decay amplitudes has the 
effect of reducing the magnitude of the $Q_6$ 
contribution to $\epsilon^\prime /\epsilon$ which one obtains
in the isospin-conserving (IC) limit.  
This is conventionally represented 
by a multiplicative factor $1-\Omega_{st}$ applied to the 
$\Delta I=1/2$ matrix element.
Explicitly, one writes \cite{BJL93}
\begin{equation}
{\epsilon^{\prime}\over \epsilon} \sim \left [ P^{(1/2)}-P^{(3/2)}\right ]
\label{sensitivity}
\end{equation}
with
\begin{eqnarray}
P^{(1/2)} & = & \sum y_i\langle Q_i\rangle_0(1-\Omega_{st}), \cr
P^{(3/2)} & = & {1\over\omega}\sum y_i\langle Q_i\rangle_2,
\label{sens2}
\end{eqnarray}
where $\omega\simeq 1/22$, the $y_i$ are the parts of the Wilson coefficients 
associated with the top quark (and hence, through the corresponding
CKM matrix elements, with direct CP-violation),
and the subscripts 0 and 2 denote the isospin of the $\pi\pi$ final state.
Note that we have dropped overall factors in Eq.~\ref{sensitivity} in
order to highlight the dependence on $\Omega_{st}$.  
In typical analyses one finds a significant
cancellation between the $\Delta I = 1/2$ and $\Delta I = 3/2$ contributions,
and therefore a non-trivial
sensitivity to $\Omega_{st}$.  

The IB correction, $\Omega_{st}$, is obtained as follows.
Writing the isospin decomposition of 
the $K^0\to\pi\pi$ decay amplitudes as
\begin{eqnarray}
A_{00} &=& \sqrt{1\over 3}A_0 {\rm e}^{i\phi_0}-\sqrt{2\over 3}A_2 {\rm e}^
{i\phi_2}, \cr
A_{+-} &=& \sqrt{1\over 3}A_0 {\rm e}^{i\phi_0}+{1\over\sqrt{6}}A_2{\rm e}^
{i\phi_2}, 
\label{isodecomp}
\end{eqnarray}
where $A_0$ and $A_2$ are, respectively, the (in general complex-valued) 
$\Delta I = 1/2$ and $\Delta I = 3/2$ amplitudes, and 
$\phi_i=\delta_i^{\pi\pi}$ are the usual  
$\pi-\pi$ scattering phases,
$\Omega_{st}$ is given by the ratio 
\begin{equation}
\Omega_{st} = {{\rm Im}\delta A_2 \over \omega{\rm Im}A_0} 
\label{omegaIB}
\end{equation}
where $\delta A_2$ is the octet 
leakage contribution to the $\Delta I = 3/2$ amplitude, and 
$\omega=({\rm Re}A_2/{\rm Re}A_0)\simeq 1/22.2$
(reflecting the $\Delta I = 1/2$ rule enhancement).
\begin{footnote}
{We follow standard phase conventions in which $A_0$ and $A_2$ are
real in the absence of CP-violation.}
\end{footnote}

The possibility that the octet leakage contribution to the matrix element
of $Q_6$ 
induced by {\bf $\delta m\not= 0$} and electric charge differences could
have a significant impact on estimates of $\epsilon^{\prime}/\epsilon$ was
first discussed in Refs.~\cite{oldamherst,oldother}. 
At leading order (LO) in the chiral expansion, the leakage
contribution to $A_2$ is saturated by $\pi^0$-$\eta$ mixing and the kinematic
effect produced by the $K^0$-$K^\pm$ mass splitting and the 
momentum-dependence of the LO weak vertices.  This leads to
the well-known LO value $\Omega_{st}=0.13$.
The difference between this LO value and the 
conventional value employed in recent analyses 
of $\epsilon^\prime /\epsilon$, 
$\Omega_{st}=0.25\pm 0.08$~\cite{oldother}, 
results from an estimate of those NLO effects mediated
by the $\eta^\prime$ through the $K\rightarrow\pi\eta^\prime$
transition and $\pi$-$\eta^\prime$ mixing, which effects
would be expected to be dominant in the large-$N_c$ limit.
In the framework of the conventional low-energy effective theory
employed in this paper (that
involving only the $\pi$, $K$ and $\eta$ degrees of freedom,
in which the $\eta^\prime$ and other higher resonances
have been integrated out),
such effects correspond to contributions to the NLO
weak LEC's (to be discussed below), although use of 
phenomenological values for the 
octet-singlet mixing angle also effectively 
incorporates NLO $\eta^\prime$-mediated
contributions proportional to the renormalised strong LEC,
$L_7^r$, of Gasser and Leutwyler~\cite{glchpt}.
As has been recently pointed out, however, other NLO contributions
might also be important.  An example is that discussed in
Ref.~\cite{emnp99}.  If one considers the effect of NLO strong
dressing on the external legs in $K\rightarrow\pi\pi$, there
is a large NLO $\eta^\prime$-induced contribution (proportional
to $L_7^r$) associated
with treating $\pi^0$-$\eta$ mixing at NLO.  This mixing contribution,
however, always occurs in the fixed combination $3L_7^r+L_8^r$
(see, for example, the expressions for the 
angles, $\hat{\theta}_1$ and $\hat{\theta}_2$,
describing NLO mixing, given in
Ref.~\cite{kmmix}).  As pointed out in Ref.~\cite{emnp99},
there is a strong numerical cancellation between the $\eta^\prime$-induced
$L_7^r$ and the scalar-resonance-induced $L_8^r$ contributions,
clearly demonstrating the importance of including NLO contributions
other than those induced by the $\eta^\prime$.
The effect of this cancellation was found, in Ref.~\cite{emnp99},
to lower 
$\Omega_{st}$ to $0.16\pm 0.03$ (thus increasing the Standard Model
prediction of $\epsilon^{\prime}/\epsilon$ by about 21\% \cite{emnp99}).  
The possibility of additional
non-$\eta^\prime$-induced NLO contributions to $\Omega_{st}$
was also recently explored in Ref.~\cite{gv99}.
The focus of this work was
on IB contributions associated 
with NLO weak LEC's, and 
the numerical results suggest the possibility of very large 
corrections to $\Omega_{st}$ associated with scalar meson exchange.
However, as we discuss below, the 
results of Ref.~\cite{gv99} suffer from important technical shortcomings
which make them numerically unreliable.  
The significance of the possible reduction in the value of $\Omega_{st}$
suggested by Refs.~\cite{emnp99,gv99} is obvious from Eqs.~\ref{sensitivity} 
and \ref{sens2}.

Chiral perturbation theory (ChPT) provides a natural framework for the 
calculation
of $\Omega_{st}$ since it ensures that all contributions 
of a given chiral order may be obtained in a 
computationally straightforward manner.  The complete
set of NLO contributions is a sum of
NLO strong LEC, one-loop, and NLO weak LEC contributions,
each of which is separately renormalisation-scale-dependent, 
divergent, and therefore unphysical.  
The sum of these contributions is, however,
necessarily finite and scale-independent.  The cancellation of both 
divergences and scale-dependence in the final result
provides a highly non-trivial check of the explicit calculations.
The set of Feynmann graphs to be evaluated is shown in
Fig.~1, where Fig.~1(a) represents the LO,
Fig.~1(c) the NLO strong LEC, Figs.~1(b) and 1~(d)-(g)
the one-loop, and Fig.~1(h) the NLO weak LEC
contributions, respectively.
In Ref.~\cite{emnp99} only the NLO contributions
associated with external line dressing 
(Figs.~ 1(b,c)) were considered, while Ref.~\cite{gv99} examined only the 
contribution of Fig.~1(h).

In Refs.~\cite{cwkm99,thesis} it was 
shown that, in the CP-even sector, the inclusion of
one-loop isospin-breaking corrections in $K\to\pi\pi$ decay amplitudes acts to 
{\it decrease} the magnitude of the octet leakage contribution to the 
$\Delta I = 3/2$ amplitude.  This decrease ensures that the $\Delta I = 1/2$
rule enhancement obtained in an 
IC analysis is accurate to better than 10\%.
In the present paper we obtain a 
complete NLO (${\cal O}(p^2 \delta m)$) determination of 
$\Omega_{st}$ in ChPT, and
show that the same qualitative situation holds 
in the CP-odd sector.
This means, according to Eq.~\ref{omegaIB}, 
that one should expect 
a further reduction of $\Omega_{st}$ beyond that associated with the NLO 
strong mixing effects studied in Ref.~\cite{emnp99}.  
The absence of reliable values for  
the CP-odd weak LEC's is {\it the} crucial stumbling block on the way to an 
accurate numerical result for $\Omega_{st}$ at NLO,
the NLO strong LEC and loop corrections being, as we will see below,
well-determined.  
In the present work we appeal to two models (the weak
deformation model, and the chiral quark model) of the
contribution of the gluonic penguin operator, $Q_6$, to the CP-odd
weak LEC's in order to probe the probable scale of the model-dependence in our
estimates of $\Omega_{st}$.  
It is important to emphasize that,
while the estimates of the weak LEC contributions to $\Omega_{st}$
are model-dependent, the one-loop and strong LEC contributions (discussed
below) are, though scale-dependent, model-independent.
As we will see, the combination of these model-independent
NLO contributions is, at typical hadronic scales,
rather large, and negative,
suggesting a significant reduction of $\Omega_{st}$ as compared
to the conventional value.

The rest of the
paper is organised as follows:  In the next section we briefly review the 
chiral Lagrangian approach to the calculation of non-leptonic kaon decay 
amplitudes and discuss the models employed for the relevant NLO weak LEC combinations.
In Section III we present our numerical results for $\Omega_{st}$.  The impact 
of our findings on
theoretical estimates for the value of $\epsilon^{\prime}/\epsilon$ in
the Standard Model are discussed and conclusions are presented in Section IV. 

\section{CP-Odd $K^0\to\pi\pi$ Decay Amplitudes in Chiral Perturbation Theory}

\begin{figure}
\centering{\
\psfig{figure=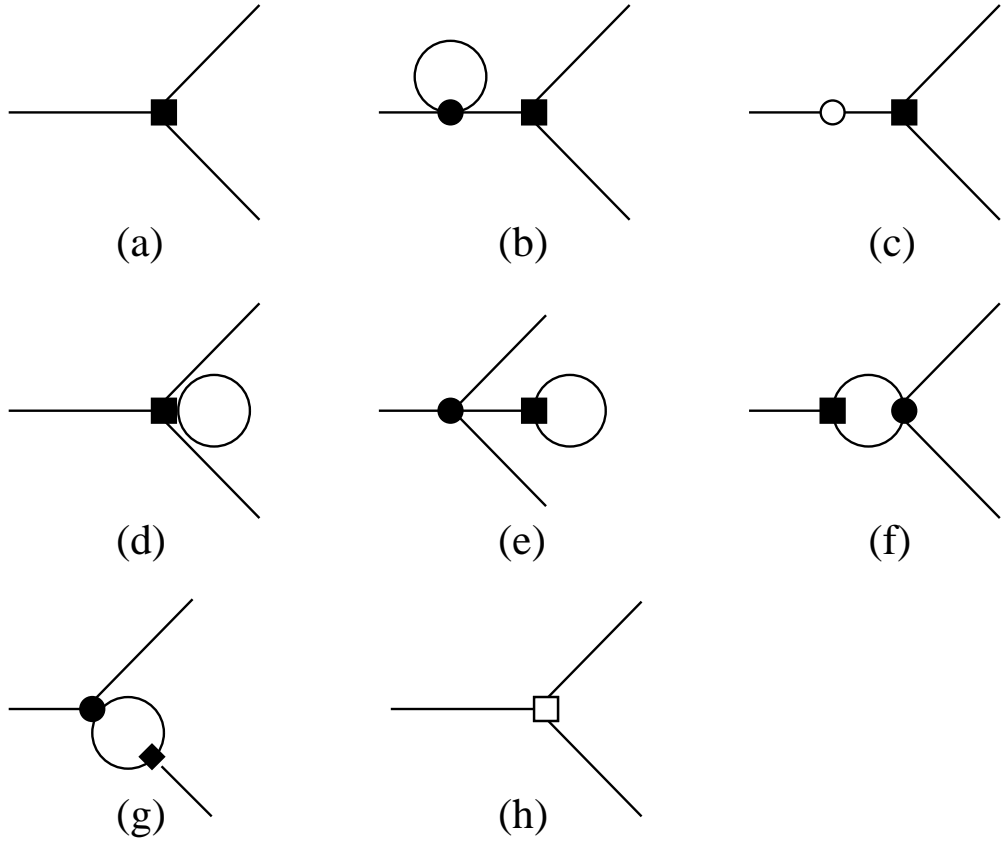}}
\vskip 0.5in
\caption{Feynman diagrams for $K\to\pi\pi$ up to ${\cal O}(p^4)$ in
the chiral expansion.  Closed circles represent ${\cal O}(p^2)$ strong
vertices,
open circles ${\cal O}(p^4)$ strong vertices, closed boxes ${\cal O}(p^2)$ weak
vertices, and open boxes ${\cal O}(p^4)$ weak vertices.  No one-line weak
tadpoles occur because, in the weak effective Lagrangian employed, they have
already been rotated away. Figures (b) and (c) should be understood
to represent collectively the strong dressing on all the external lines.}
\label{diagrams}
\end{figure}

The diagrams which have to be calculated to obtain the $K\to\pi\pi$ decay 
amplitudes to NLO are, as noted above, 
those given in Fig.~\ref{diagrams}.  
We now briefly 
review the ingredients needed for these calculations, referring the reader 
to Ref.~\cite{cwkm99,thesis,cwkm00_big} for the technical details.
The low-energy representation of the non-leptonic weak interactions is
obtained from the effective chiral Lagrangian, ${\cal L}_W$, which was
written to LO in 
Ref.~\cite{cronin}, and up to NLO in Ref.~\cite{kmw90} (or, in equivalent
reduced forms, in Refs.~\cite{ekw93,bpp98}).  
(In the CP-odd case, the gluonic penguin operator, 
which is the focus of the present work, and which,
together with the electroweak penguin operator,
dominates $\epsilon^\prime /\epsilon$ in the Standard Model,
is pure octet; we, therefore, need only the octet
components of ${\cal L}_W$.)
We work with a form of the effective weak chiral Lagrangian in which the weak
mass term appearing at LO~\cite{cronin} 
has been rotated away~\cite{kmw90}.
The LO (second order in the chiral counting) 
part of the octet Lagrangian, ${\cal L}^{(2)}_{W(8)}$, is 
thus given, in the absence of
external fields, by~\cite{cronin}
\begin{equation}
{\cal L}_{W(8)}^{(2)}=
c_2^{\pm}{\rm Tr}\left[\lambda^{\pm}\partial_{\mu}U^\dagger
\partial^{\mu}U\right]
\label{LWoctlead} 
\end{equation}
where $U=\exp({\rm i}\lambda\cdot\pi/F)$, with
$\pi^a$ the usual octet of pseudoscalar
fields, and $F$ is the pion decay constant in the chiral limit.  
The superscripts
$\pm$ label the CP-even and odd cases respectively, 
with $\lambda^+=\lambda_6$ and
$\lambda^-=\lambda_7$.  $c_2^-$ thus represents the LO 
CP-odd octet weak coupling strength.

The NLO octet weak effective chiral Lagrangian is similarly given by
either~\cite{ekw93}
\begin{equation}
{\cal L}_{W(8)}^{(4)} = {c_2^{\pm}\over F_{\pi}^2}\sum_{i=1}^{37}N_i^{\pm}
O_i(\lambda^{\pm}) ,
\label{L4nlo}
\end{equation}
or~\cite{kmw90,bpp98}  
\begin{equation}
{\cal L}_{W(8)}^{(4)} = \sum_{i=1}^{48}E_i^{\pm}\tilde{O_i}(\lambda^{\pm})
\label{L4nlokmw}
\end{equation}
where the operators which contribute to the $K\to\pi\pi$ amplitudes correspond
to $i=\{5,6,...13\}$ and $i=\{1,...5,10,...15,32,...40\}$ 
in Eqs.~\ref{L4nlo} and ~\ref{L4nlokmw}
respectively.  In quoting our results below, we will
employ the notation of Eq.~\ref{L4nlo}, and hence work with
the operator basis given by
\begin{eqnarray}
O_5(\lambda^{\pm}) & = & 
{\rm Tr}\left[\lambda^{\pm}\{S,L_{\mu}L^{\mu}\}\right] \cr
O_6(\lambda^{\pm}) & = & 
{\rm Tr}\left[\lambda^{\pm}L_{\mu}\right]{\rm Tr}\left[
SL^{\mu}\right] \cr
O_7(\lambda^{\pm}) & = & 
{\rm Tr}\left[\lambda^{\pm}S\right]{\rm Tr}\left[L_{\mu}L^{\mu}
\right] \cr
O_8(\lambda^{\pm}) & = & 
{\rm Tr}\left[\lambda^{\pm}L_{\mu}L^{\mu}\right]{\rm Tr}
\left[S\right] \cr
O_9(\lambda^{\pm}) & = & 
{\rm Tr}\left[\lambda^{\pm}\{P,L_{\mu}L^{\mu}\}\right] \cr
O_{10}(\lambda^{\pm}) & = & {\rm Tr}\left[\lambda^{\pm}S^2\right]  \cr
O_{11}(\lambda^{\pm}) & = & 
{\rm Tr}\left[\lambda^{\pm}S\right]{\rm Tr}\left[S\right] \cr
O_{12}(\lambda^{\pm}) & = & {\rm Tr}\left[\lambda^{\pm}P^2\right]  \cr
O_{13}(\lambda^{\pm}) & = & 
{\rm Tr}\left[\lambda^{\pm}P\right]{\rm Tr}\left[P\right] 
\label{ekwops}
\end{eqnarray}
where $L_{\mu} = {\rm i}U^{\dagger}\partial_{\mu}U$, 
$S=\chi^{\dagger}U+U^{\dagger}\chi$ 
and $P={\rm i}(\chi^{\dagger}U-U^{\dagger}\chi)$, 
with $\chi = 2B_0M_q$ (where $M_q$ is the quark mass matrix).
Note that in Eq.~\ref{L4nlo} the weak LEC's 
are expressed as products of factors
$c_2^-/F_{\pi}^2$ and $N_i^-$.  The latter will, henceforth, be referred to as 
reduced CP-odd NLO weak LEC's.  
(As we will see below, this reduced form has certain
advantages for estimates of $\Omega_{st}$.)

The remaining ingredient needed in order to calculate 
the diagrams of Fig.~\ref{diagrams} is the strong chiral 
Lagrangian.  We use the standard form of Gasser and Leutwyler 
given by ${\cal L}_S={\cal L}_S^{(2)}+{\cal L}_S^{(4)}+\cdots$,
where the superscripts indicate the chiral order and,
in the absence of external fields, one has~\cite{glchpt}
\begin{eqnarray}
{\cal L}_S^{(2)} &=& {F^2\over 4}{\rm Tr}[\partial_{\mu}U
\partial^{\mu}U^{\dagger}] +
{F^2\over 4}{\rm Tr}[\chi U^{\dagger}+U\chi^{\dagger}],\label{Ltwo} \\
{\cal L}_S^{(4)} &=&L_1({\rm Tr}[\partial_{\mu}U
\partial^{\mu}U^{\dagger}])^2
+L_2{\rm Tr}[\partial_{\mu}U\partial_{\nu}U^{\dagger}]
\, {\rm Tr}[\partial^{\mu}U\partial^{\nu}U^{\dagger}]
+L_3{\rm Tr}[\partial_{\mu}U\partial^{\mu}U^{\dagger}
\partial_{\nu}U\partial^{\nu}U^{\dagger}]
\nonumber \\
&&\ \
+L_4{\rm Tr}[\partial_{\mu}U\partial^{\mu}U^{\dagger}]
\, {\rm Tr}[\chi U^{\dagger}+U\chi^{\dagger}]
+L_5{\rm Tr}[\partial_{\mu}U\partial^{\mu}U^{\dagger}
(\chi U^{\dagger}+U\chi^{\dagger})] +
L_6({\rm Tr}[\chi U^{\dagger}+U\chi^{\dagger}])^2 \nonumber \\
&&\ \
+L_7({\rm Tr}[\chi U^{\dagger}-U\chi^{\dagger}])^2
+L_8{\rm Tr}[\chi U^{\dagger}\chi U^{\dagger} +
U\chi^{\dagger}U\chi^{\dagger}]
+H_2{\rm Tr}[\chi\chi^{\dagger}]\label{Lfour}\ ,
\end{eqnarray}
where $\{ L_i\}$, $F$ and $B_0$ are
the usual strong LEC's,
in the notation of Ref.~\cite{glchpt}.  Recall that, when using
dimensional regularisation, the NLO LEC's, $\{L_i\}$, 
are formally divergent and 
have a Laurent expansion in $d-4$
(where $d$ is the spacetime dimension) of the form
\begin{equation}
L_i = \Gamma_i\lambda+L_i^{r}+L_i^{(-1)}(d-4) + ...
\label{laurent}
\end{equation}
where 
\begin{equation}
\lambda = {1\over 32\pi^2}\left[\left({2\over d-4}\right)+
\gamma_E-1-{\rm ln}(4\pi)\right].
\label{lambda}
\end{equation}
The $\{L_i^r\}$ are the usual
scale-dependent renormalised versions of the $L_i$~\cite{glchpt}, 
for which we employ the 
values found 
in Ref.~\cite{ecker95}, 
while the $\{\Gamma_i\}$ are constant coefficients (frequently called
scaling coefficients) which govern the scale dependence of
the $L_i^r$.
The $\{L_i^{(-1)}\}$ contribute first to physical 
observables at next-to-next-to-leading (sixth) order (NNLO),
through one-loop graphs involving a single NLO vertex
proportional to $L_i$ and, as such, are on a similar footing
as the LEC's present in ${\cal L}_S^{(6)}$.  
(The NLO weak LEC's, $\{N_i^{\pm}\}$,
in Eq.~\ref{L4nlo}, of course, have a similar expansion.)

The {\it formal} difference between the $K^0\to\pi\pi$ vertices extracted 
from Eq.~\ref{LWoctlead} using $\lambda^+$ and $\lambda^-$ is the 
switch $c_2^+ \rightarrow {\rm i}c_2^-$.  This is also, therefore, 
the only difference 
between the corresponding LO CP-even and CP-odd decay 
amplitudes.  
Since the NLO strong LEC and one-loop contributions to these
amplitudes involve a single LO weak vertex, one readily sees, from
Fig.~1, that the substitution $c_2^+ \rightarrow {\rm i}c_2^-$
also converts the CP-even version of these contributions
into the corresponding CP-odd version.
The substitution $c_2^+ N_i^+\rightarrow {\rm i}c_2^- N_i^-$, similarly
accomplishes the CP-even$\rightarrow$ CP-odd
conversion of the weak LEC contributions, Fig. 1(h).
If one considers the {\it ratio} of the NLO contributions to
the LO contributions, therefore, the
only difference between CP-even and CP-odd 
cases for the $K^0\to\pi\pi$ amplitudes
is the difference in the numerical values and 
physical interpretation 
of the renormalised reduced weak LEC's, $N_i^r$.  
Thus, for example, the contributions to ${\rm Im}\delta A_2$
arising from the diagrams of Fig. 1(b)-(g) are immediately obtained
from the first row of Table II of Ref.~\cite{cwkm99} by multiplying the
entries by ${\rm i}c_2^-$. 
The formal contributions from the CP-odd NLO weak LEC's, 
$[A_0]_{WLEC}$ and $[\delta A_2]_{WLEC}$, arising from Fig.~\ref{diagrams}(h) 
are given in the Appendix in terms of the $N_i^-$ of Eq.~\ref{L4nlo}.  
Unlike the case of the CP-even sector, where linear combinations of octet NLO 
weak LEC's corresponding to isospin-conserving contributions were fit to the 
available $K\to\pi\pi$ and $K\to\pi\pi\pi$ data in 
Ref.~\cite{kmw91}, both the CP-odd isospin-breaking 
{\it and} isospin-conserving NLO weak LEC combinations are
unknown.  As there is not sufficient data available 
to perform fits analogous to those in the CP-even sector, 
it is necessary to resort to models to estimate the weak LEC contributions. 

In general the numerical value of the isospin-conserving combination of NLO 
weak LEC's, which appears in $[A_0]_{WLEC}$ (Eq.~\ref{weakct} 
of the appendix), 
can be determined from 
the expressions for the hadronic matrix elements at NLO 
(for which many calculations exist).
However the IB combination which appears in $[\delta A_2]_{WLEC}$ 
cannot be so determined.  
In what follows we consider three models which could be used to estimate the 
NLO weak LEC contribution to $\Omega_{st}$.  These are the weak 
deformation model (WDM) of Ref.~\cite{ekw93} 
from which direct NLO weak LEC estimates 
are available, the chiral quark model ($\chi$QM) as implemented by the Trieste 
group in Refs.~\cite{trieste95,trieste98_1,trieste_misc} for which all the 
necessary ingredients required to calculate the NLO weak LEC contributions 
to ${\rm Im}A_0$ and ${\rm Im}\delta A_2$ are readily 
available, and the ``scalar saturation model'' of Ref.~\cite{gv99},
which combines the factorisation approximation 
with the assumption of scalar meson exchange saturation of the relevant
strong LEC's.

The weak deformation model of Ref.~\cite{ekw93} proceeds from the observation
that the LO weak chiral 
Lagrangian of Eq.~\ref{LWoctlead} can be generated from 
the LO strong chiral Lagrangian of Eq.~\ref{Ltwo} by a simple
``topological deformation''.  
The model hypothesis is that this same deformation 
can be used to generate the entire $\Delta S = 1$ chiral Lagrangian.  
The WDM thus provides no information about the LO weak LEC values,
$c^\pm$, but gives explicit expressions for the reduced NLO weak LEC's,
$N_i^{\pm}$, in terms of the NLO strong LEC's, $L_i$.
For the NLO weak LEC's relevant to $K\to\pi\pi$ we have \cite{ekw93}
\begin{eqnarray}
[N_5^-]_{WDM} &=& -{3\over 2}[N_6^-]_{WDM} = -[N_7^-]_{WDM} = -L_5 \cr
[N_8^-]_{WDM} &=& 4L_4+2L_5.
\label{wdm}
\end{eqnarray}
All other NLO weak LEC's appearing in Eq.~\ref{L4nlo}
vanish in the WDM.  
Since, in these relations, the divergent parts of the $\{L_i\}$ 
do not generate the correct divergent parts of 
the $\{N_i^-\}$, one must interpret Eqs.~\ref{wdm} as
applying to the {\it renormalised} versions of the LEC's.
Moreover, since the scaling of the weak LEC's is not correctly
given by that of the strong LEC's, Eqs.~\ref{wdm} can
be taken to hold only at a single scale, which is assumed to
be a typical hadronic scale, $\mu_h$.  Assuming
resonance dominance suggests $\mu_h\sim m_{\rho}$.
Note also that, because of the scale dependence of the weak LEC's,
even though, in the WDM, the remaining $N_i^r$ vanish at the assumed
matching scale $\mu_h$, they are non-zero at other scales.
Numerical values for the IC and IB combinations of NLO weak LEC's relevant
to $K^0\to\pi\pi$ in the WDM can be found in the last column of 
Table~\ref{table1}, where we have used
conventional values for the renormalised strong
LEC's, $L_4^r$ and $L_5^r$ at 
$\mu=m_{\rho}$ taken from Ref.~\cite{ecker95}.

In another approach, the hadronic matrix elements of the four-quark 
operators relevant for 
non-leptonic kaon decay have been estimated in the $\chi$QM
by the Trieste group~\cite{trieste95,trieste98_1,trieste_misc}.  
In particular, expressions
for the gluonic penguin contribution to the LO 
weak LEC's were obtained in the second of Refs.~\cite{trieste_misc}, and 
preliminary estimates 
of the contribution to NLO weak LEC's were given in Ref.~\cite{trieste95}.  
As pointed out by the authors of Ref.~\cite{trieste95}, however,
the latter expressions contain errors and 
are not to be used \cite{bertolini_private}.
We have used the updated results of
Ref.~\cite{trieste98_1}, which give the contributions
of $Q_6$ to $A_0$ to NLO, to extract the CP-odd
weak LEC combination which enters ${\rm Im}A_0$.
To obtain the IB combination of NLO weak LEC's entering $\delta A_2$,
we have calculated the matrix element $\langle\pi^+\pi^0|Q_6|K^+\rangle =
(\sqrt{3}/2)\delta A_2$ in the $\chi$QM using the formalism described in
Ref.~\cite{trieste98_1}, together with the
corrected results for the basic ingredients
required to compute such matrix elements in the model 
given there.
The model parameters which enter these estimates
are the constituent quark mass, $M$, the vacuum quark condensate, and the 
gluon condensate.  These parameters were constrained to reproduce the 
$\Delta I = 1/2$ Rule in the CP-even sector in Ref.~\cite{trieste98_1} using
a matching scale $\Lambda_{\chi} = 0.8$ GeV (the
scale at which the scale-dependence of the chiral loops and that of the
short-distance expressions is roughly matched in the model).
The fitted parameters have the values $M = 0.2\pm0.02\;{\rm GeV}$, 
$\langle\bar{q}q\rangle = (-0.240^{+30}_{-10}\;{\rm GeV})^3$, and 
$\langle \alpha_sGG/\pi \rangle = (0.334\pm0.004\; {\rm GeV})^4$.  
The resulting values for the NLO weak LEC combinations are presented in the 
first column of Table~\ref{table1}.  Numerical estimates 
for the NLO weak LEC contributions to 
${\rm Im}A_0$ and ${\rm Im}\delta A_2$ are presented in the next 
section.

The third approach to estimating the CP-odd weak LEC's is
that of Ref.~\cite{gv99}.  In this reference, one begins with
the ``factorisation approximation'' 
for $Q_6$~{\begin{footnote}{The factorisation approximation becomes exact 
in the limit of large $N_c$.  In this limit, taking $Q_6$
for example, if one renormalises the two densities
of which $Q_6$ is a formal product, then one will also
have renormalised the four-quark operator $Q_6$.
For two
such renormalised densities, $J(x)$ and $J^\prime (y)$, at {\it different}
points $x$, $y$, one can straightforwardly construct the
low-energy representation of the product $J(x)J^\prime (y)$
using standard methods.  In general, of course, the low-energy
representation of such a product is {\it not} simply the product
of the low-energy representations of the individual densities,
but also contains seagull terms.
It turns out that, for
$Q_6$, the NLO part of this representation,
which is the part investigated by the authors of Ref.~\cite{gv99}, 
does indeed contains seagulls.
Since these diverge as $x\rightarrow y$, it is necessary to
interpret the factorisation approximation as corresponding
to an approximate low-energy representation of $Q_6$ obtained
by dropping the seagull terms in the low-energy
representation of $lim_{x\rightarrow y}J(x)J^\prime (y)$, i.e.,
to one obtained by taking simply 
the product of the low-energy representations
of the two densities.}\end{footnote}},
in which the low-energy representation of $Q_6$ is assumed
to be given by the product of the low-energy representations
of the scalar densities of which the unrenormalised
operator is formally a product.  Since the low-energy representation
of each such density is
obtained by taking the derivative of ${\cal L}_S[\chi ,\chi^\dagger ]$,
with respect to the appropriate component of $\chi$ or
$\chi^\dagger$ (treated here as external sources)~\cite{glchpt},
the model version of the low-energy
representation of $Q_6$ becomes the product of two such derivatives.
This product can be organized by chiral order.  The LO term
(second order in chiral counting), arises from the product
of the derivatives of ${\cal L}_S^{(2)}$ and ${\cal L}_S^{(4)}$,
and leads to the conventional factorisation approximation
for $c_2^-$, 
\begin{equation}
c_2^- = {G_f\over\sqrt{2}}V_{us}V_{ud}{\rm Im}C_6(16B_0^2F_{\pi}^2L_5)
\label{c2fact}
\end{equation}
where $C_6$ is the Wilson coefficient accompanying $Q_6$
in the expressions for the effective weak
Hamiltonian.{\begin{footnote}{In ChPT, $c_2^-$ is finite, and
scale-independent, whereas $L_5$ is divergent.  To make sense
of this relation one, therefore, usually assumes that 
$L_5$ is to be replaced by its renormalised value, $L_5^r(\mu)$,
evaluated at some typical hadronic scale $\mu =\mu_h$.}\end{footnote}}
At NLO, having dropped the seagull contributions associated with
the second derivative of ${\cal L}_S^{(8)}$ with respect to the 
sources, one is left with two types of terms, those involving
a product of two derivatives of ${\cal L}_S^{(4)}$, and those
involving a product of one derivative of ${\cal L}_S^{(2)}$
and one of ${\cal L}_S^{(6)}$.  
Let us denote the resulting
contributions to the factorisation approximation for the
(non-reduced) NLO weak LEC's, $E^-_i$, (as employed in Ref.~\cite{gv99})
by $\left[ E^-_i\right]_1$ and $\left[ E^-_i\right]_2$,
respectively.  The authors of Ref.~\cite{gv99} then
employ a model in which the relevant {\it renormalised}
fourth order and sixth order strong LEC's are assumed to be saturated by 
scalar resonance exchange.  There is, however, a problem with
the approach of Ref.~\cite{gv99}, associated with
the $\left[ E^-_i\right]_1$ contributions.  To understand the origin
of this problem, consider, for example, 
the results of Eq.~17 of Ref.~\cite{gv99},
translated into our notation:
\begin{equation}
\left[ E_1^-\right]_1=\left[ E_3^-\right]_1=
\left[ -E_5^-\right]_1=
{G_f\over\sqrt{2}}V_{us}V_{ud}{\rm Im}C_6(32B_0^2) L_8^2 .
\label{l8squared}
\end{equation}
As noted above, $L_8$ has a Laurent expansion of the form given by
Eq.~\ref{laurent}.
Note, first, that this means
that the contributions $\left[ E_i^-\right]_1$ of
Eq.~\ref{l8squared} begin at $O[1/(d-4)^2]$, in contrast to the
actual $E_i^-$, whose Laurent expansions begin at $O[1/(d-4)]$.
While one might plausibly ignore this discrepancy, arguing 
that only the finite parts of the expressions at some hadronic
scale are to be used in any case, a related problem remains,
even for the finite parts.  Explicitly, the fact that both
factors of $L_8$ in Eq.~\ref{l8squared} contain a $1/(d-4)$
term means that the finite part of $L_8^2$ is {\it not}
$\left[ L_8^r\right]^2$, as assumed in Ref.~\cite{gv99},
but rather $\left[ L_8^r(\mu ))\right]^2+5/(384\pi^2)L_8^{(-1)}(\mu)$,
where the explicit value of $\Gamma_8$, given in Ref.~\cite{glchpt}
has been used.
Since, as explained above, 
the $L_i^{(-1)}$ are on the same footing as the sixth order strong
LEC's which enter $\left[ E_i^-\right]_2$, the model expressions
of Ref.~\cite{gv99} for the sum 
$\left[ E_i^-\right]_1+\left[ E_i^-\right]_2$ are numerically
incomplete.  The model, moreover, provides no means of estimating
the $L_i^{(-1)}$, making it impossible to correct this defect.
In view of this problem, we conclude that, at present, it
is not possible to estimate the NLO weak CP-odd LEC's using
the factorisation approximation.

Given the problem just discussed with the numerical estimates
of Ref.~\cite{gv99}, we restrict our attention to the WDM and
$\chi$QM in obtaining estimates for the NLO weak LEC contributions
to ${\rm Im}A_0$ and ${\rm Im}\delta A_2$.  The resulting
values for $\Omega_{st}$ will be given in the next section.  Since the
$\chi$QM is a microscopic model, and the WDM is not, we will
take the value of $\Omega_{st}$ obtained using the former model
as our central value, and use the deviation from this central value
of the result obtained using the WDM as a {\it minimal}
measure of the theoretical uncertainty in our prediction
for $\Omega_{st}$ associated with the model dependence of
the weak NLO LEC's.

\section{Numerical Results}
The isospin-breaking correction to the gluonic penguin operator, 
$Q_6$, evaluated 
to ${\cal O}(p^2\delta m)$, 
can be written in terms of its LO (${\cal O}(\delta m)$) value and NLO 
corrections as
\begin{equation}
\Omega_{st}^{(2)}\left[1+R_2-R_0\right ]
\label{omegaIBnlo}
\end{equation}
with
\begin{equation}
\Omega_{st}^{(2)} = {\sqrt{2}\over 6\omega}{B_0\delta m\over (\bar{m}_K^2-
m_{\pi}^2)}
\label{omegaIBleading}
\end{equation}
where $\bar{m}_K = (m_{K^0}+m_{K^+})/2$.  (Note that the result
of Eq.~\ref{omegaIBleading}
is unambiguous and independent of the LO weak coupling $c_2^-$.)
The NLO corrections, $R_i$, are given by
\begin{equation}
R_0 = {{\rm Im}\;A_0^{(NLO,ND)}\over {\rm Im}\;A_0^{(LO)}}, \;\;\; 
R_2 = {{\rm Im}\;\delta A_2^{(NLO,ND)}\over {\rm Im}\;\delta A_2^{(LO)}}
\label{nlocorr}
\end{equation}
where the superscript $(NLO, ND)$ indicates the sum of non-dispersion NLO
contributions (involving NLO weak and strong LEC's and the non-dispersive
parts of loop graphs).  
Note that, in 
Eq.~\ref{nlocorr}, $R_2$ arises from IB effects, whereas 
$R_0$ is purely isospin-conserving.  (The IB correction
to $R_0$ would generate a contribution to $\Omega_{st}$ of 
${\cal O}\left( [\delta m]^2\right)$, 
and thus is beyond the scope of the present work.)
To separate the model-independent contributions associated with strong LEC
and loop effects (Figs.~1(b)-(g)) from those of the
model-dependent NLO weak LEC terms (Fig.~1(h)),  it
is convenient to further expand $R_i$ as
\begin{equation}
R_i = R_i^{(non-WLEC)} + R_i^{(WLEC)}
\end{equation}          
where the superscripts indicate NLO weak LEC (WLEC) and 
one-loop-plus-strong-LEC (non-WLEC) contributions respectively.
We begin our discussion with the non-WLEC contributions.  These are
model-independent and unambiguous,
albeit renormalisation-scale-dependent,
since the NLO contributions concerned all involve exactly one weak
${\cal O}(p^2)$ vertex.
The resulting overall factor of $c_2^-$ 
in the non-WLEC part of the numerator of Eq.~\ref{nlocorr} therefore cancels
with the corresponding factor in the denominator.  
This cancellation
removes all of the short-distance uncertainties (Wilson coefficients, CKM
matrix elements, etc.) contained in $c_2^-$.
In addition, because
the LO coupling strength cancels
and the strong vertices (if any) are identical
for the CP-even and CP-odd cases, diagram-by-diagram,
the non-WLEC contributions to $\Omega_{st}$ 
are `universal', that is,
they are the same for CP-even and CP-odd cases.
To evaluate these 
contributions we use 
as numerical input the values $m_{\pi} = 135\;{\rm MeV}$, $\bar{m}_K =
495\;{\rm MeV}$, $m_{\eta} = 549\; {\rm MeV}$, and 
\begin{equation}
B_0\delta m = 
\left ({m_d-m_u\over m_d+m_u}\right )m_{\pi}^2 = 5552\pm 674\; {\rm MeV}^2 
\label{deltam}
\end{equation}
(as determined by Leutwyler in Ref.~\cite{leut96}).  
With these values, we have the usual result 
$\Omega_{st}^{(2)} = 0.128\approx 0.13$.
The NLO non-WLEC contributions to $R_0$ and $R_2$ are given in 
Table~\ref{table2}.  The results are presented at 
two different renormalisation scales, $\mu=m_{\eta}$ and $\mu=m_{\rho}$,
in order to display explicitly the scale dependence
of the non-WLEC contributions.  When using the WDM 
estimate for the NLO weak LEC's, we employ
$\mu=m_{\rho}$ (consistent with the expectations of 
resonance saturation), and when using the $\chi$QM estimates,
$\mu=0.8$ GeV (the matching scale employed
in Ref.~\cite{trieste98_1} in obtaining
fits for the $\chi$QM parameters).

It is immediately apparent that the non-WLEC contributions to $R_0$
and $R_2$ both act to reduce $\Omega_{st}$ as compared to 
its LO value.  The $R_2$ non-WLEC contribution is weakly
scale dependent and, being ``universal'',
follows immediately from the corresponding
CP-even results of Ref.~\cite{cwkm99}.
Although for scales $\mu\sim m_\rho$
$R_0^{(non-WLEC)}$ is positive, and hence acts to lower
$\Omega_{st}$, the scale dependence, in this case,
is significantly stronger.  The increase in the magnitude
of ${\rm Im}A_0$ associated with the loop contributions is
what one would expect given the attractive final state
interactions (FSI) in the $I=0$ channel, and is analogous
to the $A_0$ FSI enhancement discussed previously
for the CP-even case~\cite{buras87,isgur90,kmw91}.
The effect of FSI on the CP-odd amplitudes
has also been recently discussed in 
Refs.~\cite{pallpich99,burasfsi00,pallente00}.
In Refs.~\cite{pallpich99,pallente00}
it is argued that the value of ${\rm Im}A_0$ obtained
from approaches which do not generate the final state
$\pi\pi$ $I=0,2$ phases for the $I=0,2$ $K\rightarrow\pi\pi$
amplitudes should be enhanced by FSI by a factor
of $\sim 1.55$, while the value of ${\rm Im}A_2$ should
be suppressed by the weakly repulsive $I=2$ FSI
by a factor of $\sim 0.92$.  The numerical values of
the enhancement/suppression are obtained using the
Omnes representation for the amplitude, and correspond
to the subtraction point $s=0$, for which
the ChPT representation of the amplitude is presumed
to be accurate.  It should be borne in mind that
the $I=0,2$ FSI, corresponding to Fig.~1(f), 
are already correctly included in our calculations,
up to NLO in the chiral expansion.  
That the $I=0$ FSI, for example, produce a significant
enhancement of ${\rm Im}A_0$, follows from
the known FSI enhancement of $A_0$ in the CP-even case~\cite{kmw91}
and the ``universality'' of the one-loop contributions.
One should also note that the Omnes function part of
the representation of the amplitudes does
not incorporate all of the NLO effects; some NLO effects
remain in the polynomial prefactor.  Thus
the question of interest to us, namely whether 
the {\it complete} set of NLO contributions raises or lowers
${\rm Im}A_0$ {\it relative to its LO value} (i.e., whether
$R_0$ is positive or negative) is not determined solely
by the character of the loop contributions;
it is perfectly possible, in principle,
for the NLO LEC contributions to be sufficiently negative
that the full NLO determination of $R_0$ is negative,
even in the presence of the attractive FSI phases.
For the models we have considered, this is not the case,
and the combination of non-WLEC and WLEC contributions
to $R_0$ is positive, leading to a suppression of 
$\Omega_{st}$ below its LO value.  
It is important to note that, so long as one adheres to the convention
of incorporating the effect of the $I=2$ leakage contribution
by means of a multiplicative correction factor applied to the 
contribution to the $I=0$ amplitude, 
there is an amplification effect at work
in the gluonic penguin contribution to $\epsilon^\prime /\epsilon$
associated with the NLO contributions to ${\rm Im}A_0$:
the more NLO effects {\it increase} the isospin-conserving
contribution to ${\rm Im}A_0$, the more
they simultaneously {\it decrease} $\Omega_{st}$.  Since
the contribution to $\epsilon^\prime /\epsilon$, including IB, is
proportional to the product of the isospin-conserving
contribution and the factor $1-\Omega_{st}$, both effects serve
to enhance the $Q_6$ contribution to $\epsilon^\prime /\epsilon$.

The NLO weak LEC contributions to $R_0$ and $R_2$ are
estimated using the models described in the previous section.  The numerical
results are displayed in Table~\ref{table3}.  We note first that the 
contributions to $R_0^{(WLEC)}$ and $R_2^{(WLEC)}$ in the WDM are identical, 
and hence cancel in the difference, $R_2-R_0$,
entering Eq.~\ref{omegaIBnlo}.  
In the $\chi$QM, the WLEC contributions to $\Omega_{st}$ are positive.  
Indeed the results of 
Tables~\ref{table2} and \ref{table3} show significant cancellation between 
the WLEC and non-WLEC contributions in the $\chi$QM.

The total NLO correction factor, $1+R_2-R_0$ which multiplies
$\Omega_{st}^{(2)}$, resulting from the combination of WLEC and non-WLEC 
contributions, is 
\begin{equation}
1+R_2-R_0 = \Bigg\{ {0.64\;\;\;\,(\chi{\rm QM})\atop 0.27\;\;({\rm WDM}).}
\label{results}
\end{equation}
Taking the $\chi$QM result as a central value, 
and the deviation of the WDM result from this central
value as a minimal measure of the model-dependence of our
result, we find 
the IB correction to the gluonic penguin contribution 
to $\epsilon^{\prime}/\epsilon$ to be 
\begin{equation}
\Omega_{st} = 0.08\pm0.05\pm0.01
\label{omegaIBmean}
\end{equation}
where the first error represents the uncertainty associated
with the model dependence of the NLO weak LEC's,
and the second the uncertainty in the input value of
$B_0\delta m$.  The central value in Eq.~\ref{omegaIBmean}
is significantly lower than both the conventionally-employed 
value, $0.25\pm0.08$, and the 
result of Ref.~\cite{emnp99}, $0.16\pm0.03$.  
That the relative uncertainty increases from about 30\% to 62\% is a 
reflection of the uncertain state of our knowledge of the NLO weak LEC's.  
We emphasize, however, that,
regardless of the actual value of the weak NLO LEC contributions, 
the model-independent one-loop and strong LEC contributions, which are
unambiguous, produce a significant
reduction of $\Omega_{st}$ for any plausible choice of hadronic scale.  
As such, the inclusion of the loop contributions 
is crucial to any attempt to evaluate
$\Omega_{st}$ beyond LO.
For the models
considered for the weak NLO LEC's, the net effect
is to drive $\Omega_{st}$ significantly below its LO value.

\section{Discussion and Conclusions}

As noted above, 
the significant cancellation between gluonic penguin and electroweak
penguin contributions means that
predictions for the value of $\epsilon^\prime /\epsilon$ 
in the Standard Model can depend rather sensitively on
$\Omega_{st}$.  The exact
degree of sensitivity, of course, depends on the 
relative size of these two dominant contributions, on which there is,
as of yet, no clear theoretical concensus.
In order to illustrate the impact of the decrease of
$\Omega_{st}$ from the conventional central value, $0.25$,
to $0.08\pm 0.05$, let us use the 
rough approximation to Eq.~\ref{sensitivity}
discussed in Ref.~\cite{munich}
\begin{equation}
{\epsilon^{\prime}\over \epsilon} \propto \left[ B_6(1-\Omega_{st})-
0.4B_8\right] \, 
\label{munich}
\end{equation}
(where we have dropped an overall constant multiplicative factor
irrelevant to the present discussion). 
Maintaining the constraint, $B_6>B_8$, imposed by the Munich
group~\cite{munich,jamin99}, and using the 
values $B_6=1.0\pm 0.3$ and $B_8=0.8\pm 0.2$ employed by them,
we find the results shown in Table \ref{finalvalues}.
The range of values for $B_6$, $B_8$ covered in the Table
is the same as that in Table 3 of the second of 
Refs.~\cite{jamin99}, from which the values for $\epsilon^\prime /\epsilon$
corresponding to $\Omega_{st}=0.25$ have also been taken.
All results correspond to central values of 
the input parameters $\Lambda_{\overline{MS}}^{(4)}$,
$m_s(m_c)$, $m_t$ and ${\rm Im}\lambda_t$.
>From the Table we see that the decrease in $\Omega_{st}$
corresponds to an increase in $\epsilon^\prime /\epsilon$
of between $21\%$ and $63\%$ ($40\pm 11\%$ for the central 
$B_6,B_8$ values).  The increase in the magnitude of
$\epsilon^\prime /\epsilon$ is between $2\times 10^{-4}$
and $5\times 10^{-4}$, to be compared to the current experimental
world average $(19.3\pm 2.4)\times 10^{-4}$.
The magnitude of the increase will, of course, be even larger
for models with larger values of $B_6$.

It is useful to comment in more detail on the application
of the corrections discussed above to microscopic models such
as the $\chi$QM and the extended NJL model~\cite{lund}.
Such models allow one, in principle, to compute the
corrections corresponding to the NLO weak LEC's self-consistently
within the model, as was, for example, done by the Trieste
group~\cite{trieste98_1} for ${\rm Im}A_0$.  
As pointed out
in Ref.~\cite{trieste_latest} (in the context of
the $\chi$QM), however, modifying the model predictions 
for $\epsilon^\prime /\epsilon$ obtained using
the conventional value of $\Omega_{st}$ is more complicated
than simply re-scaling the gluonic penguin
contribution to take into
account the new value of the factor $1-\Omega_{st}$.
The reason is that the conventional value of $\Omega_{st}$
(assumed to be the same for the CP-even and CP-odd cases)
enters also the determination of the CP-even amplitude
$A_2$ in the model; a change in $\Omega_{st}$ in the CP-even
sector would thus necessitate a re-fitting of the parameters
of the model.  In fact, in Ref.~\cite{trieste_latest}
it was noted that the re-fitting of parameters necessitated
by a shift in $\Omega_{st}$ would almost entirely compensate
for the effect of the shifted value of $\Omega_{st}$ in
the model determination of $\epsilon^\prime /\epsilon$.
One should, however, bear in mind the caveat that 
this observation is based on the implicit assumption that
$\Omega_{st}$ is the same in the CP-even and CP-odd sectors.
Although this is true for the non-WLEC contributions, there
is no reason to expect it to be true for the NLO weak LEC
contributions.  In fact, since these contributions correspond
to the hadronization of very different effective operators,
it would be rather surprising to find them taking on the same
values.  Fortunately one does not need to speculate idly
on this question:  in models such as the $\chi$QM
it is possible to simply compute the
NLO terms corresponding to the weak NLO LEC
contributions.  Having fitted the model parameters in the
CP-even sector, one would then obtain, self-consistently,
a determination of $\left[ \Omega_{st}\right]_{WLEC}$
for both the CP-even and CP-odd cases.  In order to 
make sure the determination of the strong IB correction
to $\epsilon^\prime /\epsilon$ is
under control, it is important to separately determine
the NLO weak LEC contributions to the leakage amplitudes
in the CP-even and CP-odd sectors. 

To summarize, we have presented a 
complete NLO calculation of the isospin-breaking 
correction to the gluonic penguin operator contribution to 
$\epsilon^{\prime}/\epsilon$, $\Omega_{st}$.  It is found that 
model-independent NLO
one-loop and strong LEC contributions 
are of the opposite sign to the LO contribution, and 
numerically large, for typical hadronic scales.
Combined with model estimates for the NLO weak LEC contributions,
we find a significant reduction
of $\Omega_{st}$ as compared to the `standard' value of $0.25\pm 0.08$.  
Our final result is
\begin{equation}
\Omega_{st} = 0.08 \pm 0.05
\label{final}
\end{equation}
(where the uncertainties associated with model-dependence and $B_0\delta m$
have been added in quadrature).
We recommend that this central value, together with, to be
conservative, even larger errors, be employed in future estimates
of $\epsilon^\prime /\epsilon$ in the Standard Model.

\acknowledgements
We thank S. Bertolini and J. Eeg for clarifying the current state
of the $\chi$QM calculations and pointing out the existence of the errors
in the earlier LEC estimates.  CEW acknowledges the support of
the United States Department of Energy under grant \#DE-FG0287ER-40365.
KM acknowledges the ongoing support of the Natural
Sciences and Engineering Research Council of Canada, and the
hospitality and support of the Special Research Centre for the Subatomic
Structure of Matter at the University of Adelaide.
 
\vfill\eject                                      

\appendix
\section*{NLO Weak LEC Contributions to $K^0\to\pi\pi$}
The octet NLO weak LEC contributions to the 
$I=0$ and $I=2$ CP-odd $K^0\to\pi\pi$ 
amplitudes are given by 
\begin{eqnarray}
\label{weakct}
\left[ A_0\right]_{WLEC} &=& -{c_2^-\over F^2}
\left({\frac{2\sqrt{6}}{F^3}}\right)(m_K^2-
m_{\pi}^2)(m_K^2 \tilde{K}_1^--m_{\pi}^2 \tilde{K}_2^-) \nonumber \\
\left[ \delta A_2\right]_{WLEC} &=&
{c_2^-\over F^2}
\left({\frac{2B_0(m_d-m_u)}{\sqrt{3}F^3}}\right)\left( m_K^2\tilde{J}_3^-
-m_{\pi}^2\tilde{J}_4^- \right) \nonumber
\end{eqnarray}  
where the $\tilde{K}_i^-$ are isospin-conserving NLO weak LEC combinations
(the CP-even analogue of $\tilde{K}_1^-$ is discussed in 
Refs.~\cite{kmw91,ekw93}),
and the $\tilde{J}_i^-$ are isospin-breaking LEC combinations whose CP-even
analogues are discussed in Refs.~\cite{cwkm99,thesis}.  In the notation of 
Ref.~\cite{ekw93} these are given by 
\begin{eqnarray}
\label{iccomb}
\tilde{K}_1^- &=& \left[N_5^{-,r}-2N_7^{-,r}+2N_8^{-,r}+N_9^{-,r}\right]
\nonumber \\
\tilde{K}_2^- &=& \left[-2N_5^{-,r}-4N_7^{-,r}-N_8^{-,r}+2N_{10}^{-,r}+
4N_{11}^{-,r}+2N_{12}^{-,r} \right] \nonumber \\
\tilde{J}_3^- &=& \left[ N^{-,r}_5+6N^{-,r}_6
-2N^{-,r}_8-N^{-,r}_9-4N^{-,r}_{10}-8N^{-,r}_{12}-12N^{-,r}_{13}\right] 
\nonumber \\
\tilde{J}_4^- &=& \left[ 2N^{-,r}_5+6N^{-,r}_6+N^{-,r}_8
-2N^{-,r}_{10}-10N^{-,r}_{12}-12N^{-,r}_{13}\right].  \nonumber
\end{eqnarray}

\begin{table}
\caption{Model estimates of the NLO weak LEC combinations appearing in 
Eq.~\ref{weakct} for the WDM and $\chi$QM as described in the 
text (with $r_{\pi} = m_{\pi}^2/m_K^2$).}\label{table1}
\begin{tabular}{lll}
&$\chi$QM ($\times 10^{-3}$)& WDM ($\times 10^{-3}$) \\ 
\tableline
$\tilde{K}_1^- - r_{\pi}\tilde{K}_2^-$ & -4.024 & -0.673 \\
$\tilde{J}_3^- - r_{\pi}\tilde{J}_4^-$ & -3.571 & 0.673 \\
\end{tabular}
\end{table}

\begin{table}
\caption{The NLO non-WLEC contributions to $R_0$ and $R_2$ at the 
renormalisation scale $\mu$.} 
\label{table2}
\begin{tabular}{lll}
$\mu$&$R_0^{(non-WLEC)}$&$R_2^{(non-WLEC)}$ \\
\tableline
$m_{\eta}$ & -0.01690 & -0.2359 \\
$m_{\rho}$ & 0.4203 & -0.3147 \\
\end{tabular}
\end{table}

\begin{table}
\caption{The NLO weak 
counterterm (WLEC) contributions to the correction factors 
as estimated in the WDM and $\chi$QM.}\label{table3}
\begin{tabular}{lll}
&$\chi$QM&WDM \\
\tableline
$R_0^{WLEC}$ & -0.231 & -0.0331 \\
$R_2^{WLEC}$ & 0.205 & -0.0331 \\
\end{tabular}
\end{table}

\begin{table}
\caption{The dependence of $\epsilon^\prime /\epsilon$ on
$\Omega_{st}$ in the Standard Model assuming, for illustrative
purposes, the central values for $\Lambda_{\overline{MS}}$,
$m_s(m_c)$, $m_t$ and ${\rm Im}\lambda_t$ as given in
Refs.~[12].  The units of $\epsilon^\prime /\epsilon$
are $10^{-4}$.  The values of $\epsilon^\prime /\epsilon$
corresponding to $\Omega_{st}=0.25$ are taken from Table
3 of the second of Refs.~[12] and the range of values
for $B_6$, $B_8$ is the same as covered by that Table.}\label{finalvalues}
\begin{tabular}{llcc}
$B_6$&$B_8$&$\epsilon^{\prime}/\epsilon\ (\Omega_{st}=0.25)$&
$\epsilon^{\prime}/\epsilon\ (\Omega_{st}=0.08\pm 0.05)$ \\
\tableline
$1.0$&$0.6$&$8.4$&$11.2\pm 0.8$ \\
$1.0$&$0.8$&$7.0$&$9.8\pm 0.8$ \\
$1.0$&$1.0$&$5.5$&$8.2\pm 0.8$ \\
\tableline
$1.3$&$0.6$&$12.8$&$16.6\pm 1.2$ \\
$1.3$&$0.8$&$11.3$&$15.1\pm 1.1$ \\
$1.3$&$1.0$&$9.9$&$13.7\pm 1.1$ \\
\end{tabular}
\end{table}


\vfill\eject
 
\vfill\eject

\end{document}